\documentstyle[twocolumn,aps,epsf]{revtex}
\newcommand{\be}{\begin{equation}}
\newcommand{\ee}{\end{equation}}
\newcommand{\bea}{\begin{eqnarray}}
\newcommand{\eea}{\end{eqnarray}}
\newcommand{\FN}[1]{$\typeout{Kommentar: #1}$}

\begin{document}
\bibliographystyle{unsrt}
\title{Localized Structures in Pattern-Forming Systems}
\author{Hermann Riecke \thanks{Department of Engineering Sciences and Applied Mathematics
Northwestern University, Evanston, IL 60208, USA. This work was partially supported 
by grants from DOE (DE-FG02-92ER14303) and NASA (NAG3-2113).}}
\maketitle

\begin{abstract}
 
 A number of mechanisms that lead to the confinement of patterns to a small
 part of a translationally symmetric pattern-forming system are reviewed: 
nonadiabatic locking
 of fronts, global coupling and conservation laws, dispersion, and coupling
 to additional slow modes $via$ gradients. Various connections with
 experimental results are made.
\end{abstract}

\FN{Ising-bloch}
 
To appear in the proceedings of the IMA-workshop 
{\it Pattern Formation in Continuous and Coupled Systems}, May 1998.

\section{Introduction}

Over the past years investigations of pattern formation have been quite successful, 
in particular investigations of spatially and temporally 
periodic patterns have reached a
mature state. In recent years there have been quite a few experimental 
observations that go beyond this framework, in which spatially localized patterns
are found, i.e. structures in which certain patterns extend only over a small
part of the spatially homogeneous system. In most cases, far away from the localized
pattern the structures are asymptotic 
to the trivial, unstructures state or to a different, periodic state.

A classic example of such localized structures are propagating pulses of 
excitation in nerve conduction systems (e.g. \cite{TyKe88,Me92}). Over the years quite a large
number of qualitatively different localized patterns has been identified in 
a variety of dissipative systems. In convection of binary mixtures one-dimensional
 wave pulses and domains of waves have been found \cite{MoFi87,BeKo90,NiAh90,Ko94}. 
 In pure-fluid convection 
 in narrow channels confined domains of large convection
 rolls have been found embedded in a pattern of rolls of smaller wavelength
 \cite{HeVi92}. Qualitatively similar states arise in Taylor vortex
 flow \cite{BaAn86} and in parametrically excited surface waves 
 in ferrofluids \cite{MaRe98}. In the Taylor system the vortices with small wavenumber 
 turn out to show additional fine-scale turbulence. In Taylor vortex flow
 of a viscoelastic fluid striking, as yet unexplained two-vortex states 
 have been observed \cite{GrSt97}. 
Solitary propagating waves have been seen in chemical systems \cite{RoEr91},
 in gas discharge systems \cite{BoPu95}, and also in 
parametrically driven surface waves \cite{LiAr96}.
Recently two observations have found particular
 interest. On the surface of vertically vibrated granular material 
 circular, solitary waves (`oscillons') arise, which due to the temporal 
 symmetry of the 
 system occur in two symmetrically related forms that can bind and form chains and other
 larger arrangements \cite{UmMe96}. In electroconvection of 
 nematic liquid crystals stable
 long and narrow domains of convection waves nucleate spontaneously from the 
 very weak noise in the 
 system \cite{DeAh96a}.
 
 \FN{nonlinear optics pulses}
 
 The wide range of qualitatively different localized structures cannot be understood
 with a single mechanism, and in fact quite a few different mechanisms appear
 to be relevant. A comprehensive review and classification of these
 mechanisms would be valuable. This goal is, however, beyond the scope
 of the present paper. Instead, it is intended to provide  
 a brief discussion of
 some of the important aspects of the mechanisms. The topic of 
 sec.\ref{s:locking} is the 
stabilization through the interaction of fast and slow spatial scales.
Localization can also occur due to a conservation law or global coupling 
(sec.\ref{s:cons}). The effect of dispersion in wave systems 
is briefly discussed in sec.\ref{s:dispersion}. Localization through the coupling to 
an additional field $via$ gradients is reviewed in some detail in sec.\ref{s:grad}.
 
\section{Locking of Fronts}
\label{s:locking}
A quite general situation in which localized states are expected to exist 
arises in bistable 
systems in which fronts can connect the two coexisting states. 
The combination of two opposite
fronts leads to a localized structure. 
The simplest description of such a situation is given by a 
single-component reaction-diffusion equation. In the context of pattern-forming systems
the translation symmetry leads to a Ginzburg-Landau equation for the complex amplitude $A$
of the pattern  
\be
\partial_t A=\partial_x^2 A + \lambda A + c|A|^2A -|A|^4A. \label{e:rd1}
\ee
This equation is valid for weak bistability, i.e. $c>0$ small. 
Eq.(\ref{e:rd1}) admits front solutions $A_f^\pm(x-x_0,t)$ located at $x_0$
and connecting the basic state $A=0$ and a patterned state $A=A_0\,\exp(iqx)$. For long
times the front is expected to approach a front without any phase winding, 
$A={\cal A}\,exp(i\psi)$
with $\psi=const.$, i.e.
the wavenumber of the localized pattern is given by the critical wavenumber
\cite{SaHo92}. 
It is therefore reasonable to 
focus on the case of $A$ real, i.e. a nonlinear diffusion equation.  
 For general parameters the fronts will propagate and one of the
two states will invade the other. For one value of the control parameter, $\lambda_0$, 
however, the fronts are
stationary. A localized state can then be obtained by a combination of these two front solutions
separated by a distance $L$. Due to the interaction between the fronts a 
stationary state will be
obtained for suitably adjusted values of $\lambda$.
 For large values of $L$ the dynamics of the
fronts can be described asymptotically by an equation for $L$ alone,
\be
\frac{dL}{dt}=2 v(\lambda) - c e^{-L/\xi}, \label{e:dLdt}
\ee
where the velocity $v$ of a single front 
vanishes at $\lambda_0$ and $\xi$ characterizes the width
of the fronts. The interaction coefficient $c$ is positive and the fronts attract each
other. Since the interaction decays with distance the attraction renders the localized state unstable. Thus, stable localized states
consisting of two fronts can only be obtained if a repulsive component arises
through additional contributions. This is possible in various ways. 

For a steady 
pattern a quite general mechanism can stabilize  localized
structures \cite{Po86}. In the asymptotic expression (\ref{e:dLdt}) the velocity 
of widely separated fronts  vanishes
only for a single value of the control parameter $\lambda$. In addition, the interaction 
between the fronts is monotonically attractive, reflecting the monotonic 
nature of the fronts. In the full equations, which capture also the underlying pattern, 
the front velocity can vanish, however, over a {\it finite} range 
of parameter values \cite{Po86} due to an interaction between the position $x_0$ of the front 
and the underlying pattern. This interaction 
is lost within the envelope equation (\ref{e:rd1}) through the introduction 
of multiple spatial scales.

By modifying the projection that leads to the solvability condition (\ref{e:dLdt}) 
to be a projection over all of space rather than
just one wavelength (as it is done to obtain (\ref{e:rd1}))
a modified evolution equation for the front position $x_0$ can be 
obtained \cite{BeSh88},
\be
\frac{dx_{0}}{dt}= v(\lambda) +  f(x_0), \label{e:dLdtlock}
\ee
where $f(x_0)$ is periodic with the periodicity of the underlying pattern. 
Due to the second term a finite locking range $\Delta \lambda$ over which the
front is stationary is obtained. It is exponentially small in 
the steepness of the front, i.e. 
$f(x_0) \sim \lambda^\nu exp(-\alpha/\sqrt{\lambda})$, as is typical for nonadiabatic terms
\cite{Po86}. It is not clear whether the prefactor of the
exponential given in \cite{BeSh88} contains all the terms of the 
relevant order. 

Due to the oscillatory character of the patterned fronts their interaction will also
be modified. No detailed analytical calculation of this appears to have been done
so far. It has, however, been studied in quite some detail numerically 
\cite{SaBr96,SaBr97b,CrRi98}.
This work is motivated by the recent observation of localized excitations (oscillons) of
parametrically excited waves in granular material \cite{UmMe96}. These excitations cover only 
a single wavelength of the pattern and due to the subharmonic response consist in alternate 
phases of the driving of a single peak or a single crater. They arise in a parameter
regime in which the transition to spatially periodic waves is subcritical and 
predominantly to square rather than stripe patterns.

Since the subharmonic instability of the surface waves arises from a Floquet 
multiplier crossing the unit circle at -1, the
small-amplitude and slow-time behavior can be described by the same type of equation as that
obtained from a real eigenvalue crossing 0, with the additional
 requirement that the resulting
equations are equivariant under flipping the sign of the amplitude of the pattern. 
The latter expresses the two equivalent states that are phase-shifted with respect to each 
other by one period of the driving. 

To capture the non-adiabatic locking in a description of this system the fast spatial scales
have to be kept in the description. Therefore one is led to the use of order-parameter
equations (e.g. \cite{NeFr93}) of the Swift-Hohenberg type. 
As expected stable loalized
states are found in one dimension \cite{SaBr96} as well as in two dimensions 
\cite{SaBr97b,CrRi98}. In 
two dimensions the stable range appears to be noticeably larger than in one 
dimension \cite{CrRi98}. 
Of course, the description of such a short localized state using the
interaction of widely separated fronts can at most give qualitative insight. However, 
the numerical results are consistent with 
the exponential scaling of the locking range expected from (\ref{e:dLdtlock}) \cite{CrRi98}.
In two dimension the locking mechanism requires that the pattern be periodic 
in both directions as it is the case with 
square and hexagonal patterns. In the experiments square patterns are observed in the relevant
regime \cite{UmMe96}. The case of localized 
structures arising from hexagonal patterns has been discussed earlier \cite{ArGo90}. 

As in the experiments various bound states of oscillons of alternating polarity (peak $vs.$ 
crater) are found \cite{CrRi98,SaBr97b}. This multitude of solutions is due to the non-monotonic interaction 
between the oscillons. Such stationary multi-hump solutions have been discussed in great detail 
in the context of homoclinic orbits in reversible systems \cite{Ch98,SaJo97}. In the simple order-parameter
model, in addition, weakly bound states of equal polarity are found as well, which are, however,
quite sensitive to noise. A more detailed analysis raises the 
question whether this localization mechanism is indeed sufficient to describe the 
experiments or whether additional mechanisms
are relevant \cite{CrRi98,TsAr97}. In a number of other phenomenological 
models oscillons have been found as well, 
but the localization mechanisms have not been identified
clearly \cite{TsAr97,VeOt98,Ro98,EgRi98}.

The same mechanism has also been invoked to explain the localized waves 
in electroconvection of nematic liquid crystals (`worms') \cite{Tu97}. There
the transition to the extended
waves is, however, most likely supercritical \cite{TrKr97} and therefore 
no fronts connecting the basic state and the extended waves exist.

\FN{Interaction of oscillatory fronts have also been studie {BODE}!!!!!!!!!!!}

\section{Global Coupling and Conserved Quantities}

\label{s:cons}
Single localized states can also be stabilized through a global coupling or the presence
of a conservation law. Consider, for example, the simple extension of (\ref{e:rd1}),
\be
\partial_{t}A=\partial_x^2 A + \lambda A + c|A|^2A -|A|^4A-\kappa A \int_{-\infty}^{\infty } |A|^{2}\,dx.  \label{e:rdglobal}
\ee
Now a single domain of $A={\cal A}$, embedded in a domain with 
$A=0$, cannot grow to an arbitrary size $L$
due to the ever increasing damping of $A$ with $L$. 
Thus, stable domains can be obtained. This 
mechanism has been studied in detail in the context of current 
filaments in semiconductors \cite{AlBo98,ScZu91} and gas discharge 
systems.

A similar mechanism arises if the bifurcating amplitude is a conserved quantity.
 A simple case is given by the 
Ginzburg-Landau equation for phase separation of a conserved quantity,
\be
\partial_t{Q}=\partial_x^2 \left\{ (\lambda+\lambda_1\,Q+
\lambda_2\,Q^2)\,Q-G\partial_x^2Q\right\}.
 \label{e:kink}
\ee
The same equation is obtained for the slow variation of small perturbations $Q$ in the 
wavenumber of a steady (non-propagating) pattern. For $\lambda <0$ the pattern is unstable
with respect to the Eckhaus instability \cite{KrZi85}. In most cases this instability does not
saturate (i.e. $\lambda_2<0$) and the solution to (\ref{e:kink}) diverges
 in finite time, indicating a phase slip
through which the total phase $\int Q\,dx$ changes. In certain situations, however, 
$\lambda_2>0$ and
the instability can saturate. This occurs if the Eckhaus instability limit is non-convex \cite{BrDe89,Ri90a}.
Perturbations then grow into domains of large and small wavenumbers, for both of which 
the pattern is stable, whereas it is unstable for the spatially averaged wavenumber.
In contrast to the fronts discussed above these fronts connect one nonlinear state to another.
Note, that despite the $4^{th}$ derivative the stationary fronts are monotonic in space. 

This mechanism is presumably at the origin of the wavenumber domains found in convection in narrow
channels \cite{HeVi92} and in Taylor vortex flow \cite{BaAn86}. It has been clearly 
identified in simulations 
of a model for parametrically driven waves where the domain with large wavenumber is 
exhibiting spatio-temporal chaos characterized by the irregular occurrences 
of (double) phase slips. In that case, (\ref{e:kink}) does not correspond to
 the phase equation derived 
directly from the basic equations using a WKB-approach, but rather 
to an equation for a 
wavenumber that is spatially and temporally aveaged over
sufficiently many phase slips \cite{GrRi96}. While the diffusive dynamics of the 
averaged wavenumber have been confirmed through 
numerical simulations \cite{GrRi96}, it is yet unclear how to derive it systematically.
 Recently, predictions for (non-chaotic) wavenumber domains have been
confirmed in parametrically excited waves in ferrofluids \cite{RaRi97,MaRe98}. 

Neither the global coupling nor the conservation law are able to stabilize
multiple domains. Instead, the domains merge with each other until only one 
domain of each state is left. The time scale for this merging grows  
exponentially in the domain size 
\cite{ScZu91}. This coarsening can, however, be suppressed if the fronts exhibit 
an oscillatory interaction as discussed above \cite{RaRi95a}.

\section{Dispersion}

\label{s:dispersion}

\FN{auch Saarloos Puls Argumente}
A different type of localization mechanism arises in dispersive wave systems, the classic
example being the nonlinear Schr\"odinger equation and its dissipative counter-part the
complex Ginzburg-Landau equation,
\be
\partial_t A=d \partial^2_X A+\lambda A+c A |A|^2 +p|A|^4A. \label{e:cgl}
\ee
The nonlinear Schr\"odinger equation, for which all coefficients in (\ref{e:cgl}) 
are imaginary and $p=0$,
arises as the weakly nonlinear description
of dispersive traveling waves in the absence of dissipation and allows localized soliton solutions. 
Their localization is due to a balance between the amplitude 
dependence of the oscillation frequency and the linear dispersion. 
Due to the symmetries of the system the solitons form a continuous four-parameter family
of solutions, characterized by 
the position, phase, amplitude ${\cal A}$ and velocity $v$ of the soliton,
\be
A={\cal A} \,{\rm sech}({\cal A}(x-vt))e^{\frac{i}{2}({\cal A}^{2}-v^{2})t+ivx}.
 \label{e:sol}
\ee
In the presence of weak dissipation or parametric forcing \cite{ElMe89}, which introduces terms
of the form $\gamma A^{*}$ (cf. \ref{e:sand}) with $A^*$ 
denoting the complex conjugate of $A$,  
the symmetries involving the amplitude, the velocity, and the phase
(in the case of forcing) are
broken and the parameters of the solitons become slow functions of time 
\cite{We85,Pi87,ThFa88,Ri96}, e.g.,
\bea
\frac{d {\cal A}}{dt}=2(\lambda-d_rv^2){\cal A}+
\frac{2}{3}(2c_r-d_r){\cal A}^3+\frac{16}{15}p_r {\cal A}^5, \label{e:sola}
\eea
where $c_r$ denotes the real part of $c$, etc. 
Stable solitary waves correspond to stable fixed points of (\ref{e:sola}) and the 
corresponding equation for $v$. Thus, 
to obtain stable localized
waves arising from a Hopf bifurcation within this framework the bifurcation to the 
extended traveling waves must be subcritical ($p_r<0$). In fact, the localized
waves exist only for parameter values for which also extended waves exist (although
possibly unstably).
The perturbed solitons of the nonlinear Schr\"odinger equation were 
invoked \cite{Pi87,ThFa88} to model the 
localized wave trains that have been observed in binary-mixture convection 
\cite{MoFi87,BeKo90,NiAh90,Ko94}. Indeed, for large dispersion the envelope of the experimentally observed wave train
looks quite similar to a typical soliton of the nonlinear Schr\"odinger equation. 
\FN{continuous deformation soliton -> front?}

In other regimes, in which dispersion is weaker,
the wave pulses are better characterized as a pair of stably bound fronts
connecting the conductive state with the nonlinear wave state. The interaction between the
fronts is affected by dispersion and the amplitude dependence
of the frequency. In the limit of weak dispersion again an evolution equation for the length $L$
of the localized state can be derived \cite{MaNe90,HaPo91}
\be
\frac{dL}{dt}=2v(\lambda) -c_{1}e^{-\frac{L}{\xi}}+\frac{c_{2}}{L}   
\label{e:dLdtdisp}
\ee
The potential for a repulsive interaction ($c_{2}>0$) arises from the last term in (\ref{e:dLdtdisp}). It is
due to the differential phase winding, which arises from the amplitude dependence of the
frequency and which leads to a gradient in the wavenumber \cite{MaNe90}. 

Recently, cases of strong dispersion have been investigated in the 
Ginzburg-Landau equation in which stable saturated localized waves arise even in regimes in which
spatially periodic waves blow up in finite time \cite{PoSt98,HoSt72}. 
Even though the cubic dissipative
term is non-saturating and no fifth-order term is present to prevent blow up, it turns out that
the amplitude dependence of the frequency can be large enough to generate strong gradients
in the wave number which in turn lead to strong dissipation $via$ the diffusion term. 

\section{Gradient-Coupling to an Additional Field}

\label{s:grad}

By introducing a second field one can obtain very robust localized structures. 
A classic case is that of two coupled reaction-diffusion equations.
This type of system has been studied in great detail and it will suffice to mention
a few keywords \cite{TyKe88,Me92}.
In the paradigmatic case one variable acts as an activator and satisfies 
a nonlinear equation with an $S$-shaped nullcline of the reaction term
while the equation for the inhibitor is   
linear. In the lumped system (no spatial dependence) one 
can then distinguish three classes of dynamics
depending on the intersection of the nullclines of the two reaction terms: 
oscillatory, excitable, and 
bistable. The type of spatial structure that is obtained depends strongly on the ratio of the
diffusion coefficients. In the excitable regime and if both coefficients are of similar size
one obtains traveling pulses that have been widely used to model in particular nerve 
conduction \cite{TyKe88}. If the diffusion of the inhibitor is fast compared to that of the
activator it spreads well ahead (and behind) of the activator and localized 
{\it stationary} structures are obtained \cite{KoKu80}.  

\FN{bistable $\rightarrow$ fronts and locking
Turing subcritical locked fronts as above ..}

Less well investigated than the reaction-diffusion systems are systems in which
 the interaction between the different modes is through the gradients of one of the fields.
Such an interaction arises naturally in secondary bifurcations off a periodic pattern.
There the amplitude of the bifurcating mode is coupled to the phase of the underlying
pattern, which is a slow mode due to the translation symmetry of the system \cite{CoIo90}.
The bifurcating amplitude depends, however, not on the phase itself but on its gradient,
the wavenumber. Two cases of localized structures described by equations of this type
have been investigated \cite{CaCa92,RiPa92,Sa92}. The gradients can also arise from 
the advective nature of the interaction \cite{TsAr97,Ri92,GrRi96}.

If a one-dimensional steady pattern undergoes a secondary Hopf bifurcation the complex
amplitude $A$ of the oscillations and the phase $\phi$ of the underlying pattern
 satisfy an equation of the form \cite{CoIo90}
\bea 
\partial_t A&=& \lambda A + d \partial_x^2 A +c |A|^2A +f \partial_x \phi \, A,
\label{e:HopfA}\\
\partial_t \phi &=& \delta \partial_x^2 \phi + 
h \partial_x |A|^2 +iw (A^*\partial_x A  -  A \partial_xA^*).
\label{e:Hopfphi}
\eea
The coefficients in (\ref{e:HopfA}) are complex while those in (\ref{e:Hopfphi}) 
are real. It turns out that 
(\ref{e:HopfA},\ref{e:Hopfphi}) have an exact localized solution of the form \cite{Sa92}
\bea
A=A_0 \, {\rm sech}(kx)e^{i \omega t+i\psi},\nonumber \\
\frac{d\phi}{dx}=Q_0+Q_1 \, {\rm sech}^2(kx),
\eea
with $d\psi /dx=B_0 \tanh (kx)$ and the six coefficients  satisfying 
certain algebraic relationships.
Numerical simulations show that this solution can in fact be stable.
An important feature of this localized solution is that through the
coupling proportional to $f$ the local growth rate of the
oscillatory mode is modified by the wavenumber of the underlying
pattern and is strongly increased inside the localized
structure while it is reduced outside. 
This localized solution appears to capture the essence of the localized 
oscillations found in square electroconvection patterns
of a nematic liquid crystal \cite{KoSa94}.

A somewhat similar situation is found for patterns undergoing a parity-breaking bifurcation,
i.e. an instability that breaks the reflection symmetry of the pattern and induces 
a drift of the pattern. Such instabilities 
can also lead to localized structures in the form of domains of traveling waves that
drift through the otherwise stationary pattern. They have been observed 
in directional solidification \cite{FlSi91}, viscous fingering \cite{CuFo93}, 
and in Taylor vortex flow \cite{WiAl92}. 

The parity-breaking instability of a steady pattern 
can be described by an equation for the real 
amplitude $A$ of the asymmetric mode that breaks the reflection symmetry and again for the
phase $\phi$ of the underlying pattern \cite{CoGo89,CoIo90,FaDo90},
\bea
\partial_t A&=&(\lambda +\lambda_1 \partial_x \phi) A -A^3 + d \partial_x^2 A + b \partial_x^2 \phi +h.o.t.\label{e:pbA}\\
\partial_t \phi & =& A.
\label{e:pbphi}
\eea
Note that these equations exhibit an inhomogeneous scaling in the small parameter measuring the distance from 
threshold, the amplitude and the space and time scales.
The existence of
localized drift waves in (\ref{e:pbA},\ref{e:pbphi}) can be seen quite easily \cite{RiPa92,CaCa92}. 
Assuming a steady solution 
($q(x)\equiv \partial_{x} \phi(x)$, $A(x)$) in a frame moving with velocity $v$ 
the wavenumber of the underlying pattern is given by 
$q=-A/v +q_\infty$ and the amplitude $A$ satisfies 
\bea
d\partial_x^2 A + &&(v-\frac{b}{v})\, \partial_x A =\nonumber \\
-\frac{d}{dA} && \left\{\frac{1}{2}(\lambda +\lambda_1 q_\infty) A^2
-\frac{\lambda_1}{3v}A^3-\frac{1}{4}A^4\right\}. \label{e:pbpot}
\eea
A homoclinic orbit in space connecting $A=0$ with itself exists if the `friction'
$v-b/v$ vanishes, yielding for the velocity $v=\pm \sqrt{b}$. 
Numerical simulations show that this state can be stable. It is noteworthy that it can only exist if it drifts: 
for $v=0$ the wavenumber $q$ would diverge. 

It should be pointed out that both secondary bifurcations discussed here are supercritical
for the spatially periodic states. Nevertheless, stable localized structures are possible. 
They correspond to homoclinic rather than heteroclinic structures in space. 

Gradient coupling can also arise in systems undergoing a primary bifurcation to a patterned
state. In a phenomenological model for surface waves in vibrated granular material a coupling
of the surface oscillation amplitude $A$ to the local averaged thickness of the layer
has been introduced leading to \cite{TsAr97} 
\bea
\partial_t A &=& \gamma A^* - (1-i\omega) A + (1+ib) \Delta A - |A|^2A - \rho A,\\
\partial_t \rho& =& \alpha \nabla \cdot \left( \rho \nabla |A|^2 \right) + \beta \Delta \rho.
\label{e:sand}
\eea
Here the gradient coupling models the expulsion of material from strongly oscillating regions.
The term involving $A^*$ arises from the parametric forcing
of the system at twice the resonant frequency of the oscillatory mode $A$.
In direct numerical simulations and by using a shooting method two-dimensional
localized structures similar
to the experimentally observed oscillons were obtained. They appear in a regime in which
square patterns arise in a subcritical bifurcation. As in the two cases discussed above,
it appears to be important that the local growth rate is enhanced inside the oscillon 
and reduced outside. As in the experiments \cite{UmMe96} 
and  other models for oscillons \cite{VeOt98,Ro98,CrRi98} bound states of localized
states with opposite polarity are found. 

In traveling-wave systems gradient coupling arises quite naturally if the waves 
advect a quantity that is dynamically relevant, i.e. evolves on a time scale comparable
to that of the bifurcating amplitude. This has been studied in quite some detail motivated
by experiments on traveling waves in binary-mixture convection 
\cite{MoFi87,BeKo90,NiAh90,Ko94}
and in electroconvection of nematic liquid crystals \cite{DeAh96a}.

In binary-mixture convection, motivated by the anomalously slow drift of the pulses,
the advection of a concentration mode by the traveling wave was 
considered \cite{Ri92}. This mode can be important since mass diffusion is very slow
in liquids. It was found that such an advection not only affects the pulse velocity
\cite{Ri96}
but can also be sufficient to localize a traveling wave structure. The equations that
were used to study this mechanism describe the evolution of the complex wave amplitude
$A$ and of a real concentration mode $C$,
\bea
\partial_t A+s \partial_x A&=&d \partial^2_x A+(\lambda +fC) A+c A |A|^2 +p|A|^4A, \label{e:ecglA}\\
\partial_t C&=&\delta \partial^2_x C -\alpha C+h \partial_x |A|^2. \label{e:ecglC} 
\eea
In general, the coefficients in (\ref{e:ecglA}) are complex while those in 
(\ref{e:ecglC}) are real. The derivation of these equations from the
Navier-Stokes equation shows that in principle quite a few additional coupling terms arise \cite{Ri96}. 

To concentrate on the localization by 
the concentration mode $C$ rather than by dispersion all coefficients in (\ref{e:ecglA})
are assumed to be real. Since the system is bistable ($c>0$, $p<0$) fronts exist that connect
the conductive state $A=0$ with the traveling-wave state $A=A_0$. Their interaction is
strongly affected by the advected field. Focussing on the effect of the advection,
evolution equations for the velocities $v_{l,t}$ of the leading and of the trailing 
front can be derived in the case of weak diffusion of $A$ and $C$, small
group velocity and weak coupling \cite{HeRi95},
\bea
v_l &=& s - \frac{\gamma}{|v_l|} + sgn(v_l) \rho, \label{e:vl}\\
v_t &=& s-\frac{\gamma}{|v_t|} + 
2\gamma \frac{e^{-\alpha L/|v_l|}}{|v_l|} - sgn(v_t) \rho,\label{e:vt} 
\eea  
\FN{L Gleichung}
where $\rho$ measures the control parameter $\lambda$, $\gamma$ is
proportional to the coupling $h$, and $L$ is the (time-dependent) length of the 
pulse. Since $C$ decays over distances much larger than $A$ 
the attractive interaction between the fronts that was obtained in (\ref{e:dLdt})
is negligible in this regime. Eqs.(\ref{e:vl},\ref{e:vt}) show that
the sign of the interaction mediated by $C$ depends on the direction of propagation
of the pulse. It is repulsive only if the whole pulse drifts opposite
to the linear group velocity (for $\gamma s>0$). This can be understood with simple arguments 
considering the effect of $C$ on the local growth rate of $A$ and its consequence
for the velocity of the respective fronts. An essential ingredient is that $|C|$ is
smaller at the trailing front than at the leading front \cite{HeRi95}.

It turns out that the advected field can even lead to localized structures if the
initial bifurcation is supercritical ($c<0$). Somewhat similar to the case of the 
parity-breaking bifurcation, for $d=0$ and $\alpha=0$ 
the coupled equations (\ref{e:ecglA},\ref{e:ecglC}) can
be reduced to a single equation for $C$ that has
the form of a particle in a (cubic) potential \cite{PiRiun},
\bea
\frac{1}{2}(s-v) \frac{\delta}{h}\partial_x^2 C 
+\nonumber \\
\left( \frac{\delta}{h}+\frac{v}{2h}(s-v)-
\left(\frac{2v\delta}{h^2}+\frac{\delta}{h}\right)C-
\frac{\delta^2}{h^2}\partial_xC\right)\partial_xC\nonumber \\
= -\frac{d}{dC} \left\{ \frac{v}{2h}C^2-\frac{v}{3h}\left(\frac{v}{h}+1\right)C^3\right\}.\label{e:twpuls}
\eea
Again the velocity $v$ of the pulse is an eigenvalue and
is determined by the condition that the `work' done
by the `friction'  vanishes over the homoclinic orbit that connects $C=0$ with itself.
Since the friction is nonlinear in this system the velocity has to be determined numerically.
Direct numerical simulation of (\ref{e:ecglA},\ref{e:ecglC}) (with $d>0$) shows that 
these traveling-wave pulses can be stable \cite{Ri92}. As in the parity-breaking case (\ref{e:pbpot})
it appears to be crucial that the pulse drift ($v\ne0$). 

The advection of a slow mode by traveling waves appears also to be relevant to understand
recent observations of localized waves (`worms') in electroconvection of nematic
liquid crystals \cite{DeAh96a}. Due to the anisotropy of this system the
initial Hopf bifurcation leads to the competition of waves traveling in four 
symmetrically related directions that are oblique to the preferred direction 
characterized by the director of the liquid crystal. In analogy to zig-zag patterns
these waves may be termed left- and right-zigs and -zags, respectively. The worms
are made up of right-traveling zigs and zags and drift slowly to the left (or vice versa).
A surprising aspect of the system is that the initial bifurcation to the periodic waves 
is supercritical \cite{TrKr97}, but the worms are nucleated well below that Hopf bifurcation 
\cite{BiAh98}. The usual coupled Ginzburg-Landau equations for
the two participating wave amplitudes alone are therefore insufficient to describe the worms.
\begin{figure}[ht]
\centerline{\epsfxsize=4.in{\epsfbox{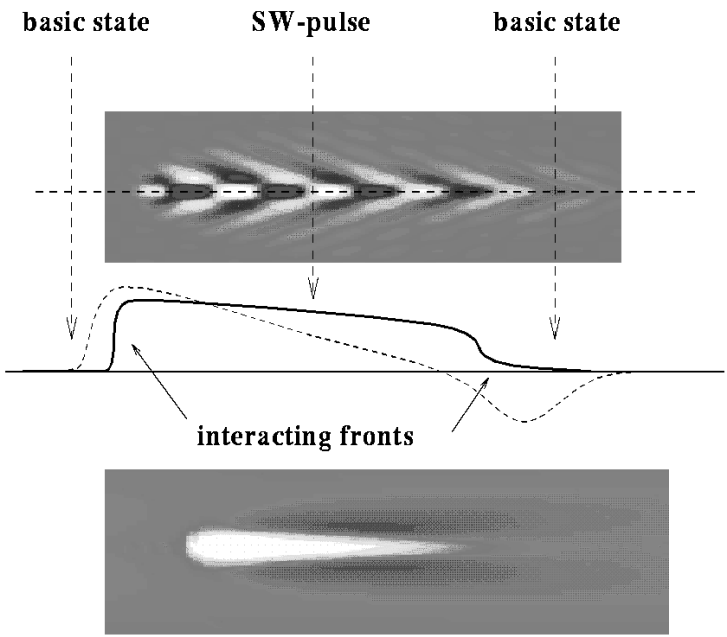}}
}
\caption{Zig- and zag-wave (top) and $C$-field (bottom) in numerical simulations
of a worm. The sketch in the center
indicates schematically the amplitude of the standing-wave pulse (solid line)
and of the $C$-field (averaged over $y$, dashed line).}
\label{f:worm}
\end{figure}
An extension that includes the advection of a slow mode similar
to the concentration mode in binary-mixture convection has been considered \cite{RiGr98},
\bea
\partial_t A&= &- {\bf u}_{A}\cdot \nabla A + \mu A + 
b_x \partial_x^2 A + b_y \partial_y^2 A + 2a\partial_{xy}^2A\label{e:cglA}\\
& &  + f C A + c|A|^2A + g|B|^2A, \nonumber\\
\partial_t B &=& - {\bf u}_{B} \cdot \nabla B + \mu B + 
b_x \partial_x^2 B + b_y \partial_y^2 B - 2a\partial_{xy}^2B\label{e:cglB}\\
& &  + f C B + c|B|^2B + g|A|^2B, \nonumber\\
\partial_t C &=& \delta \partial_x^2 C - \alpha C + 
{\bf h}_{A} \cdot \nabla |A|^2 
+ {\bf h}_{B} \cdot \nabla |B|^2. \label{e:cglC}
\eea
Here $A$ and $B$ are the amplitudes for the right-traveling zig- and zag-waves, respectively.
Within these equations the localization of the worms can be understood to arise from
the combination 
of two different mechanisms. In a one-dimensional reduction in the $y$-direction
transverse to the worm
(\ref{e:cglA}-\ref{e:cglC}) reduce to the equations for two counterpropagating waves
each advecting $C$ in opposite directions. Numerical simulations show that again a localized
structure can exist stably already below the Hopf bifurcation although that bifurcation 
is supercritical ($c<0$). Similar to the traveling-wave pulse this 
standing-wave pulse is homoclinic in space, but 
in contrast to the traveling-wave pulses no analytic description like (\ref{e:twpuls})
is available as yet. 

Given the coexistence of the standing-wave pulses (not of the extended waves) with the 
basic, non-convective state there exist also fronts that connect the basic state at 
$x=\pm \infty$ with the standing-wave pulse (see fig.\ref{f:worm}). In analogy to the 
interaction between fronts discussed in the case of binary-mixture convection 
(\ref{e:vl},\ref{e:vt}) one may expect that these fronts form a stable worm if the
worm drifts opposite to the $x$-component of the linear group velocity of the two wave
components. Indeed, the experimentally observed worms show this behavior \cite{DeAh96a}.

\section{Conclusion}

Quite a few experimentally observed localized structures in dissipative systems
can be understood qualitatively with the mechanisms discussed in this paper.
With respect to quantitative comparisons the results are somewhat limited, yet. 
Given the variety of different structures and mechanisms it would be of great
interest to condense them into paradigmatic cases, which will depend on the
symmetries of the underlying pattern (steady, traveling, oscillatory,\ldots) and 
the symmetries of the coupling to additional slow modes if present. Another relevant
distinction will be whether the localized structures are homoclinic or heteroclinic
in space. While for some of the presented mechanisms
analytical insight has been gained in limiting cases this is not the case for all of them.
For instance, a systematic treatment of localization through nonadiabatic effects,
which are quite general for steady structures and for standing waves, would be valuable.
In contrast to the exponentially small interaction between localized structures
the formation of localized structures by such an interaction between 
{\em fronts of patterns} has not been treated much.


 
\end{document}